# Decoupled symmetries of phase coherence and pairing in Eu-doped infinite-layer nickelates

Minyi Xu[1,2], Xiaofang Fu[1], Yusen Feng[1], Junlong Wu[1], Shengkun Chen[2], Zeyou Li[3], Peng Li[2], Fu-Chun Zhang[4], Guangming Zhang[5,6] and Liang Qiao[1,2*]

[1]School of Physics, University of Electronic Science and Technology of China, Chengdu 611731, China;

[2]State Key Laboratory of Electronic Thin Films and Integrated Devices, University of Electronic Science and Technology of China, Chengdu 611731, China;

[3]Department of Physics, Blackett Laboratory, Imperial College London, London SW7 2AZ, United Kingdom.

[4]Kavli Institute for Theoretical Sciences, University of Chinese Academy of Sciences, Beijing 100190, China.

[5]State Key Laboratory of Low Dimensional Quantum Physics, Department of Physics, Tsinghua University, Beijing 100084, China;

[6]School of Physical Science and Technology, ShanghaiTech University, Shanghai 201210, China;

[*]Corresponding author: L. Q. (E-mail: liang.qiao@uestc.edu.cn)

## Abstract

Superconductivity requires both electron pairing and global phase coherence, which jointly govern macroscopic superconducting transport. However, the anisotropy of the superconducting state is routinely interpreted as evidence of pairing symmetry without distinguishing their contributions, raising the question of whether pairing and phase coherence can exhibit distinct angular symmetries and which dominates the observed anisotropy. Here we show that the normal-state strange metallicity selectively imprints its anisotropy on phase stiffness rather than pairing amplitude, thereby decoupling their in-plane symmetries. By developing continuous angle-resolved pulsed transport in infinite-layer $Sm_{0.95-x}Ca_{0.05}Eu_xNiO_2$, we independently resolve a twofold phase stiffness and a predominantly fourfold depairing response. Specifically, the onset temperature of superconductivity ($T_{c,onset}$) is nearly isotropic, whereas the zero-resistance temperature ($T_{c,0}$) exhibits pronounced twofold anisotropy, locating the emergent anisotropy in global phase coherence rather than local pairing. Furthermore, the linear-in-$T$ scattering coefficient ($A_1$) shares the principal axes of phase stiffness and is linearly anticorrelated with it, whereas no analogous correspondence exists with depairing. Our work establishes a brand-new paradigm for investigating the symmetry of the complex superconducting order parameter, clarifying the intrinsic correlation among pairing, phase coherence, and strange metallicity.

## Introduction

Superconductivity requires both Cooper pairing, characterized by the pairing amplitude $|\Delta|$, and global phase coherence, sustained by the phase stiffness $\rho_s$[1,2]. The former binds electrons into correlated pairs and opens a superconducting gap. Owing to crystal symmetry[3], pairing interactions[4], or orbital selectivity[5], pairing amplitude can exhibit prominent angular dependence, which reflects the microscopic pairing interactions. In parallel, phase coherence enables the long-range ordering of Cooper pairs, which sustains dissipationless superflow. Under the influence of band dispersion[6], quantum geometry[7], or mesoscopic Josephson coupling[8], the in-plane phase stiffness can become highly anisotropic, thereby dictating the angular superfluid response. Consequently, anisotropy in superconducting transport conflates contributions from both pairing amplitude and phase coherence. However, such anisotropy is routinely taken as evidence of gap structure alone[9], leaving the disentanglement of these two contributions unresolved (Extended Data Table 1).

In conventional superconductors, pairing and phase ordering emerge simultaneously at $T_c$ with aligned or compatible symmetries, because a large phase stiffness far exceeds pairing strength[1]. By contrast, in strongly correlated unconventional superconductors, pair formation can precede phase coherence[10], where $T_c$ is limited primarily by weak phase stiffness rather than pairing[11]. This separation raises a fundamental question: in principle, can symmetry decoupling occur between pairing and phase coherence? Related ideas have been explored in pair-density-wave states[12,13], stripe-ordered systems[14,15], and quasi-one-dimensional superconductors[16]. Nevertheless, independently resolving the angular profiles of pairing strength and phase stiffness within the same superconducting state is crucial yet remains experimentally challenging.

The emergence of superconductivity from an anisotropic metallic state adds another dimension to this question. In unconventional superconductors, strange metallicity, characterized by linear-in-$T$ resistivity, is ubiquitous above the superconducting phase[17], pointing to an intimate link between anomalous normal-state scattering and superconductivity[18]. Reported correlations connect anomalous scattering with pairing[19]; the linear-resistivity coefficient $A_1$ with $T_c$[20,21]; strange-metal transport with superfluid density[22]. However, existing discussions largely treat superconductivity as a single entity, lacking quantitative comparisons among $\rho_s$, $|\Delta|$, and $A_1$. More importantly, when the strange metal itself is anisotropic, whether pairing and phase coherence inherit its symmetry remains a pending target.

Infinite-layer nickelates provide a compelling platform for testing whether pairing and phase coherence encode distinct rotational symmetries within an anisotropic strange-metal state[23]. Although nickelates share the square-planar $d^9$ motif of cuprates, their distinct multiband electronic structure provides a critical testbed for elucidating the universal mechanisms of unconventional superconductivity[24]. Evidence for strong phase fluctuations and anisotropic phase stiffness underscores the importance of disentangling the respective contributions of pairing and collective coherence[24,25]. Meanwhile, reported linear-in-$T$ resistivity and electronic nematicity raise the question of how anomalous normal-state transport governs the angular transport of superconductivity[25,26]. Together, these features motivate us to investigate how pairing stability and global phase ordering emerge from an anisotropic electronic fluid.

Disentangling pairing and phase coherence within a single device remains experimentally challenging. Combined Josephson and quasiparticle tunnelling spatially resolve superfluid density and the pair-breaking gap[28], but stringent surface requirements hinder its application to metastable infinite-layer nickelates[24,29]. Conventional continuous-current transport encounters heating and vortex-mediated dissipation before reaching the depairing limit[25,27]. Here we develop a continuously angle-resolved pulsed-transport technique using externally synchronized microsecond-short pulses[25], enabling separate resolution of phase-ordering and depairing responses in the same device. In $Sm_{0.95-x}Ca_{0.05}Eu_xNiO_2$, we uncovered a pronounced twofold phase-stiffness anisotropy coexisting with a predominantly fourfold depairing response. Moreover, we find that the $T$-linear resistivity coefficient $A_1$ presents the same twofold anisotropy and shares the same principal axis as phase stiffness. A linear anticorrelation between phase stiffness and $A_1$ is further quantified, without an analogous correspondence to depairing. Our findings establish rotational-symmetry decoupling between phase coherence and pairing, suggesting that the normal-state strange metal selectively imprints its anisotropy on phase coherence rather than pairing. Macroscopic superconducting anisotropy therefore need not reflect the underlying pairing symmetry.

# Results

## Separating phase stiffness and pairing

The interplay between rotational symmetry breaking and superconductivity has been widely explored to elucidate the nature of unconventional superconducting order parameters[30,31]. However, macroscopic superconducting anisotropy alone cannot establish whether rotational symmetry breaking resides in the $|\Delta|$, $\rho_s$, or both[1,32]. Figure 1a–c outlines three representative limiting scenarios. In a non-nematic unconventional superconductor, exemplified by an ideal d-wave state, both the pairing amplitude and phase-stiffness tensor retain the fourfold symmetry of the lattice (Fig. 1a)[30]. In a nematic superconductor, rotational symmetry breaking enters the superconducting state itself, rendering both the pairing response and, generically, phase stiffness twofold (Fig. 1b). This gap-centred picture underlies the conventional interpretation of nematic superconductivity[33]. A third possibility is a symmetry-decoupled scenario, in which the pairing amplitude remains predominantly fourfold whereas the phase stiffness becomes twofold (Fig. 1c). Here we identify such sector-selective symmetry decoupling, demonstrating that rotational symmetry need not be inherited uniformly across the superconducting condensate.

Phase stiffness and the finite-momentum depairing can be distinguished through low- and high-current regimes of pulsed current-voltage characteristics (Fig. 1d–f). At low currents, where current-induced suppression of pairing is negligible, dissipation is governed predominantly by vortex–antivortex unbinding (Fig. 1d). The resulting nonlinear relation $V \propto I^{\alpha}$ provides a transport probe of phase ordering. The Berezinskii–Kosterlitz–Thouless transition (BKT) criterion ($\alpha$=3) defines a temperature scale $T_{BKT}$, which is associated with the phase stiffness through the universal-jump relation (Fig. 1f)[25,34–36]. Higher currents instead impart finite centre-of-mass momentum to Cooper pairs and progressively suppress the condensate (Fig. 1e) [37,38]. At the depairing limit, the current-carrying condensate becomes unstable and switches onto the normal-state branch (Fig. 1i). Short microsecond pulses limit thermal accumulation, helping to approach this boundary before heating causes premature switching.

To enable continuous angle-resolved measurements of the above-mentioned characteristics, we developed a vector-current pulsed-transport method with an angular resolution of $\Delta\phi \sim 0.1°$[25,27,39]. Microsecond pulses were used to limit thermal accumulation, which can obscure the intrinsic depairing boundary. Exploiting the vector nature of current density, two orthogonal short-pulsed d.c. current components, $J_x$=$J_0\sin\phi$ and $J_y$=$J_0\cos\phi$, were independently controlled to synthesize a net in-plane current density $\boldsymbol{J}(\phi)$=$J_x\hat{x}$+$J_y\hat{y}$. By independently tuning the amplitudes of these two orthogonal pulses, the direction of $\boldsymbol{J}(\phi)$ can be continuously rotated from 0 to 360°. As a result, the simultaneously recorded voltage components, $V_x$ and $V_y$, were then projected parallel to $\boldsymbol{J}(\phi)$, yielding the longitudinal voltage $V_L$= $V_x\sin\phi$ + $V_y\cos\phi$.

To our knowledge, the only spatial comparison of both the pair-breaking gap and superfluid density was achieved by scanning Josephson tunnelling microscopy in $FeTe_{0.55}Se_{0.45}$, where the superfluid was found to be strongly inhomogeneous and independent of the pairing[28]. A symmetry-resolved comparison of superfluid and pairing has remained unavailable (Extended Data Table 1). Our technique addresses this gap by measuring the phase coherence and pairing stability on the same device.

## Distinct phase-ordering and depairing symmetries

We first determined the angular dependence of the phase coherence. In a two-dimensional superconductor, global phase coherence is established through the Berezinskii-Kosterlitz-Thouless (BKT) transition, where current-induced vortex-antivortex unbinding produces a nonlinear current-voltage response, $V \propto I^{\alpha(T,\phi)}$[35]. The nonlinear $V \propto I^{\alpha(T,\phi)}$ is measured as a function of temperature and in-plane direction $\phi$ (Fig. 2a; see more in Extended Data Fig. 4). Figure 2c shows the extracted exponent $\alpha(\phi)$ at selected values of $T$, plotted as a colour map in the $T$–$\phi$ plane. Below 20K, $\alpha(\phi)$ develops pronounced anisotropy, with its maximum near $\phi$=180°, indicating the strongest phase ordering at 180°. We then extracted $T_{BKT}(\phi)$ using the standard ($\alpha$=3) criterion[34]. The resulting $T_{BKT}(\phi)$ displays a clear twofold anisotropy, identifying a C2 global phase-ordering boundary in $Sm_{0.75}Ca_{0.05}Eu_{0.2}NiO_2$ (Fig. 2e and Supplementary Fig. 2). This assignment is independently supported by vortex-dynamics analysis, where the vortex activation energy and the phase stiffness extracted from thermally-assisted vortex motion show the same C2 anisotropy and principal axis (Extended Data Fig. 5)[25,40].

The observed C2 phase stiffness is not simply a normal-state resistance anisotropy rewritten in superconducting variables. Normal-state resistance is a dissipative transport coefficient, whereas phase stiffness is the elastic energy of the condensate against phase twisting. Its consistent appearance in $T_{BKT}(\phi)$ and vortex-activation analysis therefore identifies a genuine C2 phase-coherence sector. For a low-carrier-density two-dimensional superconductor, this anisotropic stiffness sets a directional phase-ordering boundary[1].

We next determined the angular dependence of the high-current depairing strength. Unlike conventional critical currents associated with vortex dynamics, the depairing current marks the intrinsic upper current boundary of the superconducting state and is widely regarded as a pairing-amplitude-related measure of superconducting robustness. In this regime, the imposed supercurrent gives Cooper pairs a finite centre-of-mass momentum. As the current increases, this pair momentum acts as a pair-breaking perturbation, reducing the stability of the condensate until the superconducting pairing can no longer be sustained[38,41]. In the Ginzburg–Landau theory, depairing current can be written as

$$J_d = \frac{4e\hbar}{3\sqrt{3}m^*}\frac{|\Psi_0|^2}{\xi}$$

where $m^*$ is the effective mass, $|\Psi_0|$ is the zero-current order-parameter amplitude and $\xi$ is the coherence length. To approach this intrinsic upper limit boundary, we used the angle-resolved pulsed-transport measurement described above. Microsecond-short pulses drove the condensate towards the pair-breaking limit while minimizing Joule heating. In the current-voltage ($I$-$V$) curve, $I_c$ was identified from the sharp resistive switching in the pulsed $I$-$V$ characteristics, namely, the peak in d$V$/d$I$ (Fig. 2b,d). Figure 2f presents the angular dependence of $I_c$. Notably, in stark contrast to the C2 phase stiffness, $I_c(\phi)$ exhibits a dominant fourfold modulation, with maxima near $\phi$=45° and minima near $\phi$=90°, establishing a near-C4 depairing boundary on the same device and angular coordinate (Supplementary Note 4 and Fig. S8).

The depairing-limited nature of the critical current is supported by two observations[37]. First, immediately above $I_c$, the differential resistance rapidly reaches a current-independent value comparable to the normal-state resistance, indicating direct switching to the normal state rather than an extended low-resistance vortex-flow regime (Fig. S4). Second, $I_c(T)$ follows the Ginzburg-Landau (GL) depairing-current scaling, $I_c(T) \approx I_{c,0}[1-(T/T_c)^2]^{3/2}$, for all measured angles, supporting a common depairing-limited, rather than vortex- or heating-limited, origin of the angular critical current (Fig. S5). Here, $I_{c,0}$ is the zero-temperature intercept.

To quantify the separation between angular phase stiffness and pair breaking, we decomposed each angular response, $X(\phi)=J_s(\phi)$ or $I_c(\phi)$, into isotropic, twofold, and fourfold harmonics using $X(\phi)=\eta_0+\eta_2\cos[2(\phi-\phi_2)]+\eta_4\cos[4(\phi-\phi_4)]$. The harmonic weights place phase ordering in the twofold-dominated regime and depairing predominantly in the fourfold regime (Fig. 2g). This distinction reframes the interpretation of angular superconducting transport. We prove that a twofold superconducting transport signal can reflect phase-coherence anisotropy while the high-current depairing response retains a different rotational structure. Assigning transport anisotropy to the pairing state therefore requires identifying the superconducting scale that controls the measurement.

## Strange-metal dissipation and phase coherence

Having established the disentanglement between the angular symmetries of phase coherence and pairing strength, we now discuss the fundamental question of the origin of transport anisotropy. This question is invariably tied to a compelling puzzle: how the anisotropic strange-metal state evolves into superconductivity, and whether it is tied to pairing formation, phase coherence, or both.

The first comparison is between different superconducting transition temperatures. Figure 3d shows the angular dependence of the superconducting onset temperature $T_{c,onset}$, the zero-resistance temperature $T_{c,0}$ defined by the 1%$R_N$ criterion, and the BKT temperature $T_{BKT}$ (see Extended Data Fig. 6 for details). Three key features stand out. First, $T_{c,onset}$ is nearly isotropic, showing that the initial appearance of superconducting fluctuations is weakly angle-dependent. Second, $T_{c,0}$ exhibits a pronounced anisotropy, indicating that the establishment of dissipationless charge transport depends strongly on the current-flow direction. Third, the global phase-coherence temperature $T_{BKT}$ follows the same angular dependence and principal axis as $T_{c,0}$.

As such, the direction of maximum $T_{c,0}$, denoted as $\phi_T$, therefore identifies the direction with the most robust superconducting phase ordering. The difference between nearly isotropic $T_{c,onset}$ and anisotropic $T_{c,0}$, together with the axis alignment between $T_{BKT}$ and $T_{c,0}$, suggests that the angular anisotropy is amplified during the global phase coherence rather than the onset of pairing correlations. This behaviour places strong constraints on exotic scenarios involving directional phase ordering or electronically self-organized superconducting textures, and is naturally consistent with the symmetry decoupling between the pairing and phase coherence established above.

In addition to the anisotropic superconducting transition temperatures, angle-resolved $R(T,\phi)$ curves reveal an anisotropic normal-state response before the superconducting transition (Fig. 3a). The angle-resolved $R$-$T$ curves contain an extended high-temperature regime with nearly linear-in-$T$ resistance, followed by a weak resistive upturn on approaching the superconducting transition. To characterize this normal-state electronic fluid, we fitted the temperature-dependent resistance using $R(T)=R_0+AT^n$. Figure 3b shows the extracted exponent n($T,\phi$) remains close to n=1 over an extended regime, identifying a strange-metal-like transport with an anisotropic low-$T$ boundary[26].

To quantify this strange-metal response, we further fitted the high-temperature linear regime using $R(T)=R_0+A_1T$, where $A_1$ measures the linear-in-$T$ dissipative strength. The resulting $A_1(\phi)$ exhibits a robust twofold modulation, with its minimum along $\phi\approx180°$ (Fig. 3c). Notably, the minimum of $A_1$ coincides with the maximum of $T_{c,0}$ and $T_{BKT}$. Thus, the direction with the weakest strange-metal dissipation is the direction where superconducting phase coherence is most robust. This correlation indicates that stronger linear-in-$T$ dissipation is antagonistic to global phase ordering in $Sm_{0.75}Ca_{0.05}Eu_{0.2}NiO_2$.

A scaling comparison further quantifies this relationship. After normalization, $T_{c,0}(\phi)$, $T_{c,onset}(\phi)$, and $T_{BKT}(\phi)$ follow a negative trend against $A_1(\phi)$ (Fig. 3e). Figure 4a-c summarises these angular responses: the phase stiffness $J_s(T_{BKT}^-)$, the extrapolated depairing current $I_{c,0}$, and the normalized inverse normal-state $T$-linear dissipation, $(A_1^{-1})_{norm}=A_1^{min}/A_1(\phi)$. $A_1^{min}/A_1(\phi)$ follows the same twofold axis and angular envelope as $J_s(T_{BKT}^-)$, showing that the direction with weaker normal-state $T$-linear dissipation is also the direction with stronger phase coherence. By contrast, $I_{c,0}$ retains an approximately fourfold angular structure. Direct comparisons likewise reveal a strong anticorrelation between $A_1$ and $J_s(T_{BKT}^-)$, without an analogous relationship to $I_{c,0}$ (Fig. 4e–f). These results show that the anisotropic normal-state electronic fluid is coupled primarily to the phase-coherence sector, rather than uniformly reshaping the entire superconducting order.

## Energetic basis of symmetry decoupling

The different angular relationships to $A_1$ motivate an energy-scale interpretation of the symmetry separation. In conventional weak-coupling BCS theory, pair formation and phase coherence emerge simultaneously[1,42]. To evaluate this in the context of other superconductors, we estimate an effective Fermi temperature, $T_F$, from the zero-temperature superfluid stiffness, $\rho_{s0}$, and compare $T_{c,onset}$ with $T_F$ on a Uemura-type logarithmic plot for representative materials (Fig. 4h, and Extended Data Table 2)[43,44]. Conventional BCS superconductors, such as Nb, Al, Sn and Zn, reside in the large-$T_F$, small-$T_c/T_F$ limit, whereas cuprates, FeSe-based superconductors and other unconventional superconductors exhibit a much larger $T_c/T_F$ ratio of about 0.01-0.125. $Sm_{0.95-x}Ca_{0.05}Eu_xNiO_2$ falls within this low-stiffness regime, yielding $T_c/T_F$ values from 0.049 to 0.098 depending on in-plane orientation.

This location strongly suggests that superconductivity in $Sm_{0.95-x}Ca_{0.05}Eu_xNiO_2$ departs from the conventional weak electron-phonon coupling and is associated with strong electron correlations, indicating a BCS–Bose-Einstein condensation (BEC) crossover scenario[45]. More importantly, it provides the energy-scale basis for the observed symmetry decoupling: a lower phase-ordering energy scale is highly susceptible to anisotropic normal-state electronic dissipation, whereas the larger pair-breaking energy remains unperturbed as it is dictated primarily by the local robustness of Cooper pairs.

As shown in Figure 4d, the symmetry decoupling places a strong constraint on the microscopic physics of superconductivity in $Sm_{0.95-x}Ca_{0.05}Eu_xNiO_2$. The key implication is that the superconducting state contains at least two energy scales with different sensitivity to the anisotropic strange-metal environment. At low current, Cooper pairs are not strongly perturbed, and transport is governed by the establishment of global phase coherence[34,46]. As the current increases, the imposed superfluid momentum approaches the depairing limit, and the superconducting boundary becomes controlled by the pair-breaking energy scale[37]. Consequently, the evolution of the angular response across the $I$-$T$ map shows that different current regimes probe different superconducting energy scales: low-drive transport reflects collective phase rigidity, whereas high-drive transport accesses local pair robustness.

# Discussion

Finally, we discuss possible microscopic origins of the symmetry decoupling. Three observations support a quasi-one-dimensional, stripe-like electronic texture for $Sm_{0.95-x}Ca_{0.05}Eu_xNiO_2$ (Fig. 4g). First, phase stiffness and depairing have different angular symmetries. Second, $T_{c,onset}(\phi)$ is nearly isotropic, whereas $T_{c,0}(\phi)$ is strongly anisotropic, locating the anisotropy during the establishment of global phase coherence rather than the onset of pairing correlations. Third, weaker strange-metal dissipation coincides with stronger phase stiffness, indicating that the anisotropic normal-state electronic fluid selectively influences the phase-coherence boundary. This hierarchy is compatible with a stripe-like or quasi-one-dimensional superconducting texture: local Cooper pairing can remain comparatively robust and retain a higher symmetry, whereas global phase coherence is established through directionally preferred channels. Electronic texture could therefore separate pair formation from long-range phase locking, producing spatially modulated coherence without requiring a twofold pairing.

This scenario may explain the relatively low $T_c$ of infinite-layer nickelates compared with cuprates. Our results support a substantial role for phase stiffness in determining $T_c$. In an anisotropic 2D superconducting condensate, the relevant $T_c$ is set not only by the larger stiffness, but by the geometric mean of the phase stiffness, $T_c \propto T_\theta \propto (\rho_x \rho_y)^{1/2}$. A quasi-one-dimensional superconducting texture therefore imposes a severe bottleneck: strong longitudinal rigidity cannot compensate for weak transverse coupling between neighbouring stripes. Consequently, enhanced pairing along one direction need not produce a proportionate increase in $T_c$ when the transverse phase stiffness remains small. This is closely analogous to stripe-ordered cuprates, where one-dimensional correlations can support local pairing, while weak inter-stripe or interlayer Josephson coupling delays the establishment of global phase coherence, thus limiting the $T_c$[14,15,47]. In this sense, the low $T_c$ of infinite-layer nickelates may thus reflect inefficient establishment of global coherence despite robust local pairing.

It is worth further exploring other possible origins, including mesoscale superconducting inhomogeneity or pair-density wave[28,32,48]. Tests using controlled strain, electrostatic or chemical doping, and disorder engineering could determine whether the decoupled amplitude and phase sectors arise from intrinsic multicomponent superconductivity, finite-momentum or pair-density-wave tendencies, or from extrinsic granular/percolative superconductivity. Also, pressure-dependent angle-resolved pulsed transport on freestanding infinite-layer nickelate offers a further test[49], motivated by the pressure enhancement of $T_c$ reported by Hwang and co-workers[50]. Within our scenario, pressure could strengthen transverse Josephson coupling as well as local pairing, increasing the geometric-mean stiffness and relieving the phase-coherence bottleneck. If so, the twofold anisotropy of $T_{BKT}$ and $T_{c,0}$ should weaken under pressure, whereas the fourfold depairing scale may remain tied to the underlying gap structure. Complementary microwave or terahertz measurements of anisotropic phase stiffness, together with spatial probes of charge, spin, oxygen, or hydrogen textures, could further constrain the microscopic origin. In addition, as symmetry-decoupled coherence-pairing processes are important in understanding the exotic physics of unconventional superconductors, our results provide a platform for further study of the universal theory of unconventional superconductivity.

**Fig. 1 | Principles of angular transport for symmetry-decoupled superconductivity.**

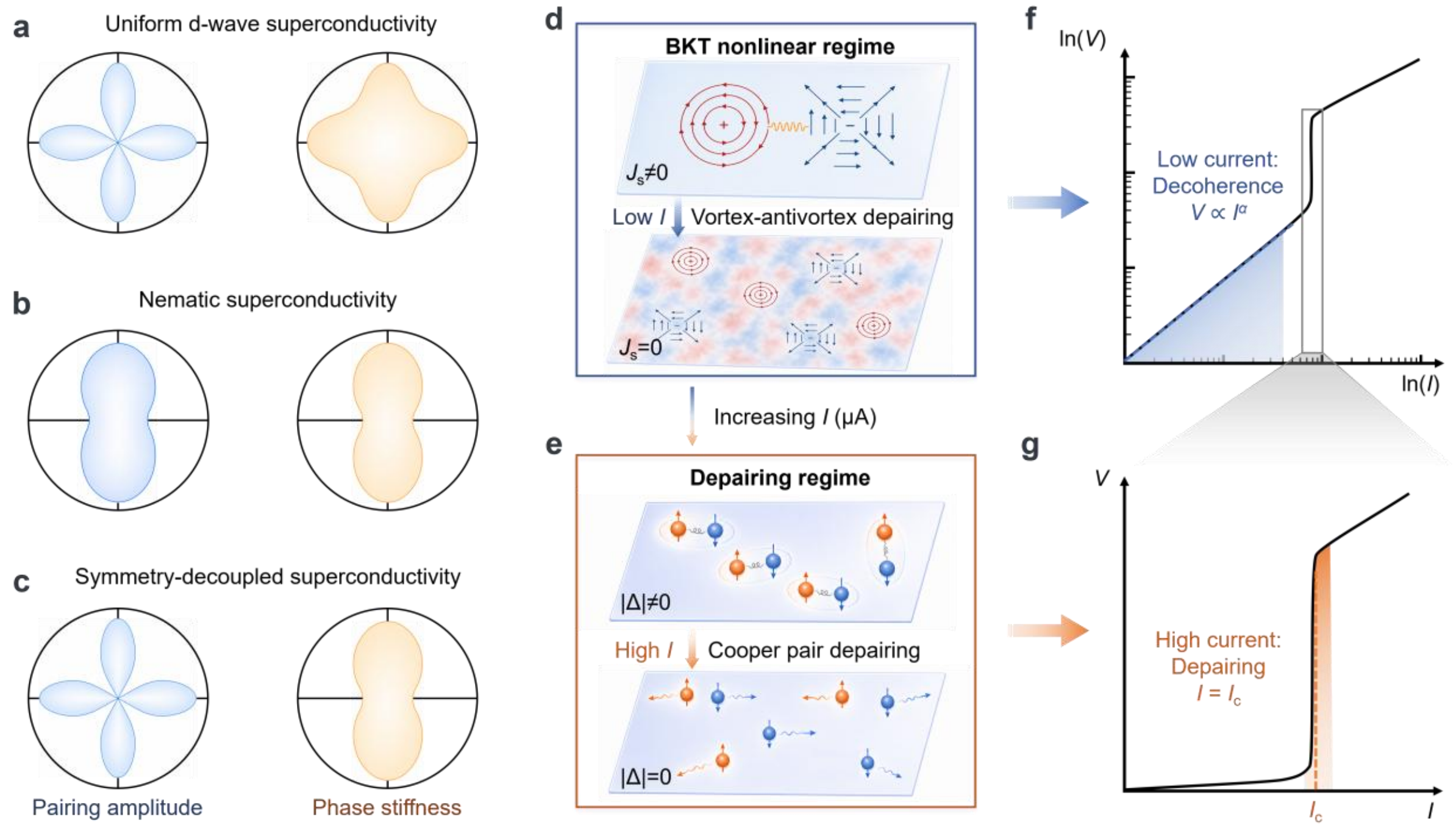

**a–c,** Schematic relations between the pairing-amplitude sector (blue) and the phase-stiffness tensor (orange). **a,** Tetragonal non-nematic $d$-wave state, in which both sectors retain fourfold symmetry. **b,** Nematic superconducting state, in which both sectors acquire a leading twofold component. **c,** Symmetry-decoupled superconducting state, where the pairing amplitude remains nearly fourfold whereas the phase stiffness becomes twofold. **d,** Schematic diagram of phase decoherence driven by vortex-antivortex unbinding in the low-current nonlinear I-V regime. The corresponding readout in **(f)** is a nonlinear voltage response $V \propto I^{\alpha}$, from which the BKT criterion ($\alpha$=3) defines the phase-ordering scale. **e,** Schematic diagram of current-driven pair-breaking in high-current regime. A large supercurrent gives Cooper pairs finite centre-of-mass momentum until Cooper pairs break and switch to a resistive state. The corresponding readout in **(g)** is the critical threshold $I_c$, which corresponds to the depairing scale.

**Fig. 2 | Twofold phase coherence and predominantly fourfold depairing in $Sm_{0.75}Ca_{0.05}Eu_{0.20}NiO_2$.**

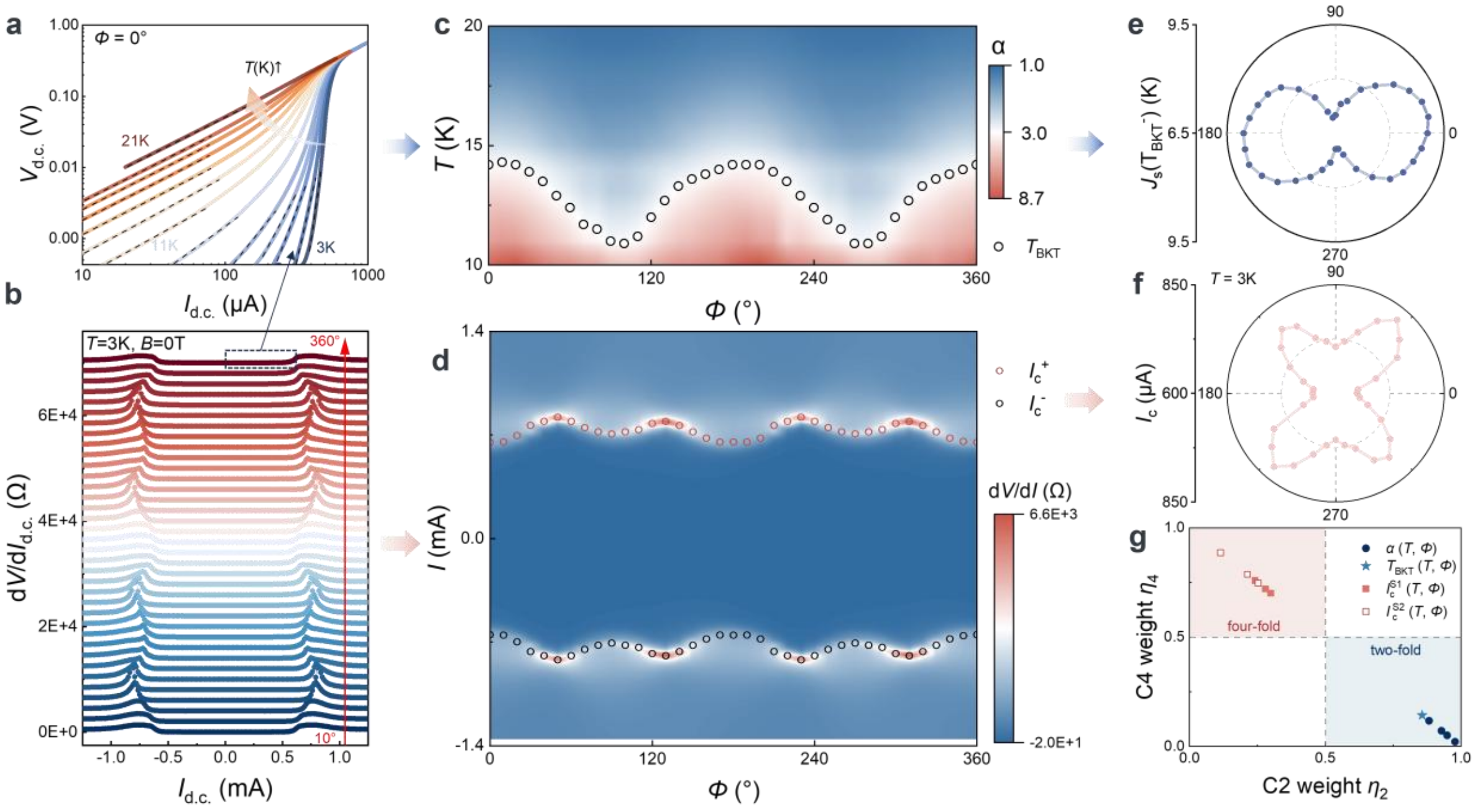

**a,** Representative $I$–$V$ characteristics measured at $\phi$=0° between 3 K and 21 K on logarithmic axes. Dashed black lines are fits to $V \propto I^{\alpha}$. **b,** Differential resistance $dV_L/dI$ as a function of bias current at 3 K for various $\phi$ separated by 10°. Successive curves are vertically offset for clarity. **c,** Colour map of the nonlinear exponent $\alpha(T,\phi)$. Open circles mark the transport-defined BKT temperature scale determined by $\alpha$=3; the extracted $T_{BKT}(\phi)$ is shown in **e**. **d**, Colour map of $dV_L/dI$ as a function of $I$ and $\phi$ at 3 K. Red and black open circles mark $I_c^+$ and $I_c^-$, extracted from the positive- and negative-bias switching peaks in $dV/dI$ and separating the central low-$dV/dI$ region from the high-current resistive branches. **e**, Angular dependence of $J_s(T_{BKT}^-)$. **f**, Angular dependence of the reciprocal critical-current component $I_{c,even} = (|I_c^+| + |I_c^-|)/2$, which is dominated by fourfold symmetry. **g**, Normalized C2 and C4 harmonic weights extracted from $\alpha(T,\phi)$, $T_{BKT}(\phi)$, and the temperature-dependent critical currents of Samples 1 and 2. The phase-ordering observables lie in the $C_2$-dominated regime, whereas the critical-current data are predominantly $C_4$. The temperature-dependent angular response of Sample 2 is shown in Fig. S8.

**Fig. 3 | Angular anticorrelation between normal-state dissipation and superconducting phase coherence.**

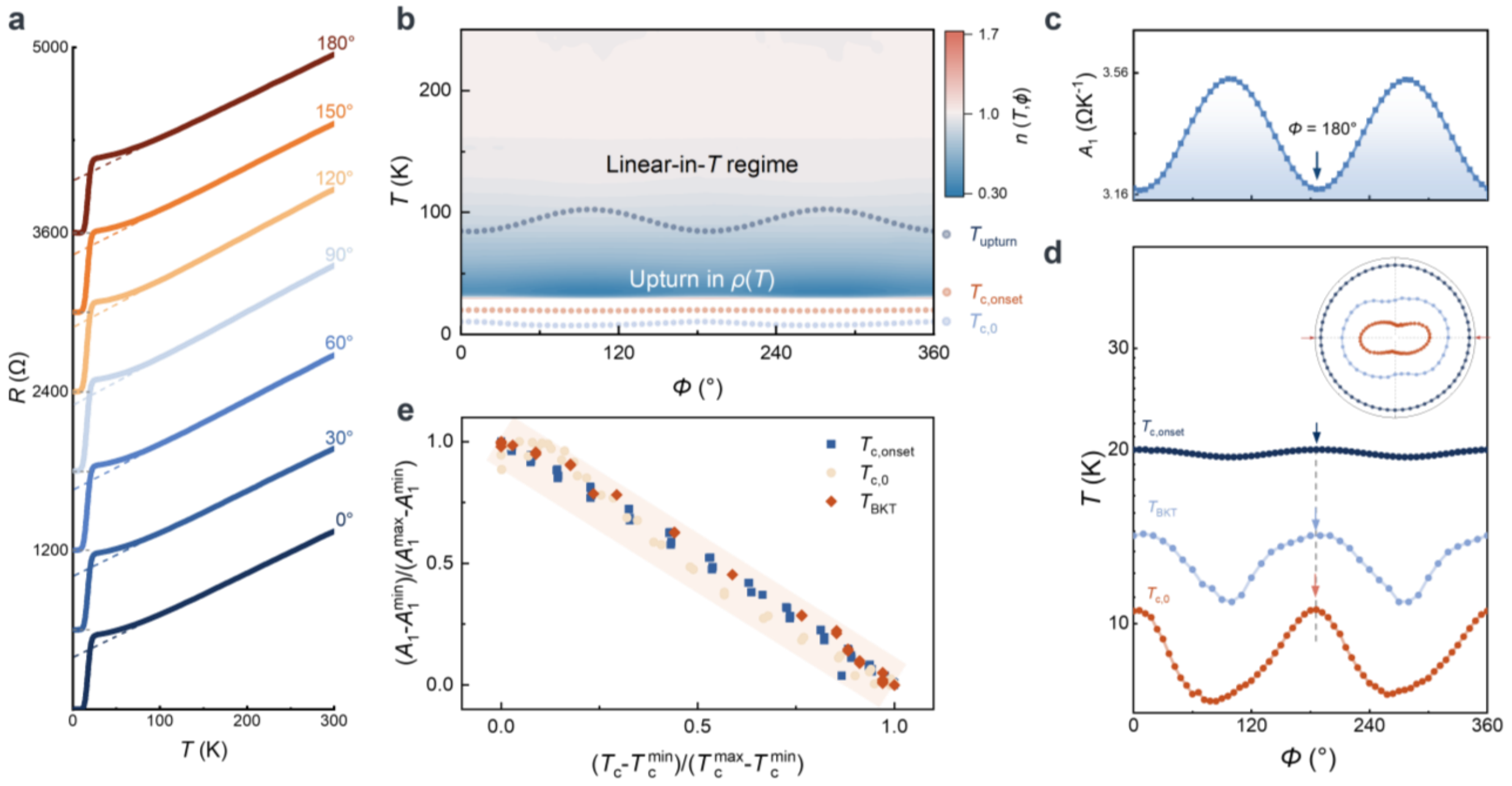

**a**, Temperature dependence of the longitudinal resistance $R(T,\phi)$ over the full-range temperature for representative $\phi$. **b**, Colour map of the resistance exponent $n(T,\phi)$, with $n \approx 1$ identifying the linear-in-$T$ regime. Dark-blue circles mark $T_{upturn}$ ($\phi$), below which $R(T,\phi)$ deviates upward from the extrapolated high-temperature linear-in-$T$ resistance. Orange and light-blue circles denote the nearly isotropic superconducting onset temperature $T_{c,onset}$ and the anisotropic zero-resistance temperature $T_{c,0}$, respectively. **c**, Angular dependence of the linear-in-$T$ coefficient $A_1(\phi)$, obtained by fitting the high-temperature linear-resistance regime to $R(T,\phi)=R_0(\phi)+A_1(\phi)T$. **d**, Angular dependence of $T_{c,onset}$, $T_{c,0}$, and $T_{BKT}$. Whereas $T_{c,onset}$ is nearly isotropic, $T_{c,0}$, and $T_{BKT}$ exhibit aligned twofold modulations. **e**, Correlation between the normalized linear-in-$T$ coefficient $A_1(\phi)$ and the normalized superconducting transition temperature scales $T_{c,onset}$, $T_{c,0}$, and $T_{BKT}$. Each point corresponds to one in-plane current direction. The negative slopes reveal an angular anticorrelation between normal-state linear-in-$T$ dissipation and superconducting phase ordering.

**Fig. 4 | Phase-stiffness-limited superconductivity and quasi-stripe scenario.**

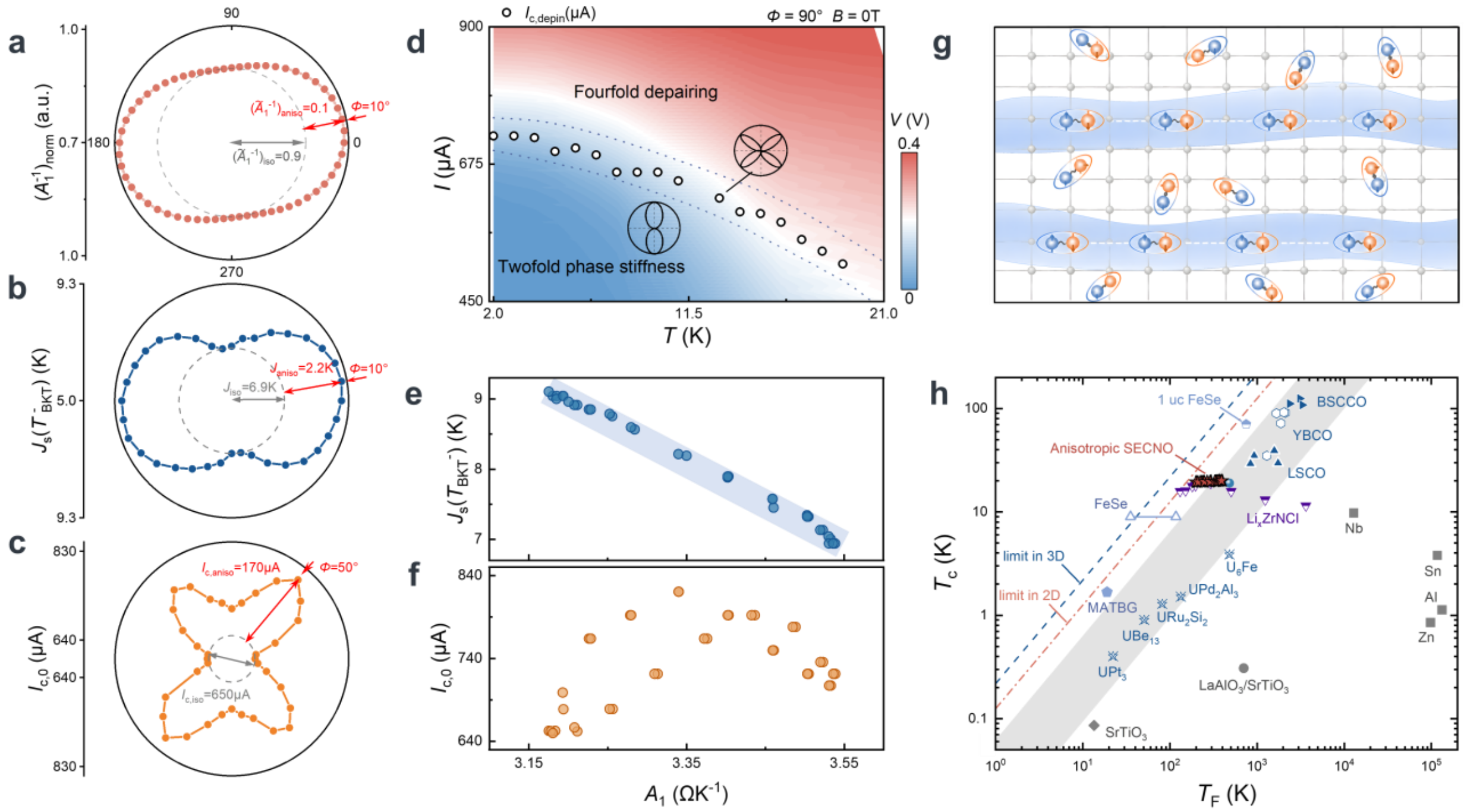

**a**, Angular dependence of the normalized inverse $T$-linear resistivity coefficient, $A_{1,\min}/A_1(\phi)$, where $A_{1,\min}$ is the minimum of $A_1$ over the measured angles. **b**, BKT-derived phase-stiffness scale, $J_s(T_{BKT}^-)=2k_BT_{BKT}/\pi$, estimated immediately below the $T_{BKT}$ and exhibiting pronounced twofold anisotropy. **c**, Angular dependence of the Bardeen-extrapolated zero-temperature critical-current scale, $I_{c,0}(\phi)$, exhibiting a predominantly fourfold symmetry. This scale is extracted by fitting the reciprocal critical current to $I_c^{even}(T, \phi)= I_{c,0}(\phi)[1-(T/T_c)^2]^{3/2}$, with $I_c^{even}=(|I_c^+|+|I_c^-|)/2$. **d**, Schematic $I$–$T$ regime diagram illustrating the crossover from a twofold phase-stiffness-limited response to a fourfold depairing-dominated response. **e**, $J_s(T_{BKT}^-)$ plotted against $A_1$, showing an approximately linear anticorrelation. **f**, Comparison between $I_{c,0}$ and $A_1$, showing no clear correlation over the measured $\phi$ range. **g**, Proposed physical picture of symmetry-decoupled superconductivity. **h**, Uemura-type $T_c$-$T_F$ plot comparing anisotropic $Sm_{0.75}Ca_{0.05}Eu_{0.20}NiO_2$ with representative superconductors. The position of $Sm_{0.75}Ca_{0.05}Eu_{0.20}NiO_2$ indicates a low-stiffness superconducting regime close to the BCS-BEC crossover scale, where phase fluctuations are important.